\documentstyle[sprocl,epsfig,psfig]{article}

\def\al{\alpha}

\def\be{\begin{equation}}
\def\ee{\end{equation}}
\def\bea{\begin{eqnarray}}
\def\eea{\end{eqnarray}}

\def\beq{\begin{equation}}
\def\eeq{\end{equation}}
\def\beqar{\begin{eqnarray}}
\def\eeqar{\end{eqnarray}}
\def\barr#1{\begin{array}{#1}}
\def\earr{\end{array}}
\def\bfi{\begin{figure}}
\def\efi{\end{figure}}
\def\btab{\begin{table}}
\def\etab{\end{table}}
\def\bce{\begin{center}}
\def\ece{\end{center}}

\def\text{\textstyle}

\def\al{\alpha}
\def\be{\beta}
\def\ga{\gamma}
\def\de{\delta}

\def\Ga{\Gamma}
\def\De{\Delta}

\def\refeq#1{\mbox{eq.~(\ref{#1})}}
\def\reffi#1{\mbox{Fig.~\ref{#1}}}

\def\refta#1{\mbox{Tab.~\ref{#1}}}

\newcommand{\GeV}{\unskip\,\mathrm{GeV}}
\newcommand{\MeV}{\unskip\,\mathrm{MeV}}
\newcommand{\TeV}{\unskip\,\mathrm{TeV}}

\def\mathswitchr#1{\relax\ifmmode{\mathrm{#1}}\else$\mathrm{#1}$\fi}

\newcommand{\PW}{\mathswitchr W}
\newcommand{\PZ}{\mathswitchr Z}

\newcommand{\PH}{\mathswitchr H}

\newcommand{\Ph}{\mathswitchr h}

\newcommand{\Pb}{\mathswitchr b}

\newcommand{\Pt}{\mathswitchr t}

\def\mathswitch#1{\relax\ifmmode#1\else$#1$\fi}

\newcommand{\MW}{\mathswitch {M_\PW}}

\newcommand{\MZ}{\mathswitch {M_\PZ}}
\newcommand{\MH}{\mathswitch {M_\PH}}

\newcommand{\Mb}{\mathswitch {m_\Pb}}
\newcommand{\Mt}{\mathswitch {m_\Pt}}
\newcommand{\scrs}{\scriptscriptstyle}
\newcommand{\sw}{\mathswitch {s_{\scrs\PW}}}

\newcommand{\cw}{\mathswitch {c_{\scrs\PW}}}
\newcommand{\sweff}{\sin^2 \theta_{\mathrm{eff}}}
\newcommand{\sweffsub}{\sin^2 \theta_{\mathrm{eff, subtr}}}

\newcommand{\MWsub}{M_{\PW, \mathrm{subtr}}}

\newcommand{\GF}{\mathswitch {G_\mu}}

\newcommand{\lsim}
{\;\raisebox{-.3em}{$\stackrel{\displaystyle <}{\sim}$}\;}
\newcommand{\gsim}
{\;\raisebox{-.3em}{$\stackrel{\displaystyle >}{\sim}$}\;}

\newcommand{\alps}{\alpha_{\mathrm s}}

\newcommand{\ses}{self-en\-er\-gies}
\newcommand{\fea}{{\em FeynArts}}

\newcommand{\two}{{\em TwoCalc}}
\newcommand{\fh}{{\em FeynHiggs}}

\renewcommand{\Re}{\mathop{\mathrm{Re}}}

\newcommand{\mst}{m_{\tilde{t}}}

\newcommand{\mste}{m_{\tilde{t}_1}}
\newcommand{\mstz}{m_{\tilde{t}_2}}

\newcommand{\Mtlr}{M_{t}^{LR}}

\newcommand{\msq}{m_{\tilde{q}}}

\newcommand{\oaas}{{\cal O}(\alpha\alpha_s)}
\newcommand{\cp}{{\cal CP}}

\newcommand{\onel}{one-loop}

\newcommand{\mh}{m_\Ph}
\newcommand{\mH}{m_\PH}

\newcommand{\Stop}{\tilde{t}}

\newcommand{\Sbot}{\tilde{b}}

\newcommand{\tst}{\theta_{\tilde{t}}}

\newcommand{\tsf}{\theta\kern-.20em_{\tilde{f}}}
\newcommand{\tsfp}{\theta\kern-.20em_{\tilde{f}\prime}}
\newcommand{\tsq}{\theta\kern-.15em_{\tilde{q}}}

\newcommand{\VL}{\left( \begin{array}{c}}
\newcommand{\VR}{\end{array} \right)}
\newcommand{\ML}{\left( \begin{array}{cc}}
\newcommand{\MLd}{\left( \begin{array}{ccc}}
\newcommand{\MLv}{\left( \begin{array}{cccc}}
\newcommand{\MR}{\end{array} \right)}

\newcommand{\Tb}{\tan \beta\hspace{1mm}}

\newcommand{\CTb}{\cot \beta\hspace{1mm}}

\newcommand{\gev}{\,\, \mathrm{GeV}}

\newcommand{\BC}{\begin{center}}
\newcommand{\EC}{\end{center}}
\newcommand{\BE}{\begin{equation}}
\newcommand{\EE}{\end{equation}}
\newcommand{\BEA}{\begin{eqnarray}}
\newcommand{\BEAnn}{\begin{eqnarray*}}
\newcommand{\EEA}{\end{eqnarray}}
\newcommand{\EEAnn}{\end{eqnarray*}}

\newcommand{\id}{{\rm 1\kern-.12em
\rule{0.3pt}{1.5ex}\raisebox{0.0ex}{\rule{0.1em}{0.3pt}}}}

\def\draftdate{\relax}
\def\mda{\relax}
\def\mua{\relax}
\def\mla{\relax}
\def\draft{
\def\thtystars{******************************}
\def\sixtystars{\thtystars\thtystars}
\typeout{}
\typeout{\sixtystars**}
\typeout{* Draft mode!
         For final version remove \protect\draft\space in source file
*}
\typeout{\sixtystars**}
\typeout{}
\def\draftdate{\today}
\def\mua{\marginpar[\boldmath\hfil$\uparrow$]%
                   {\boldmath$\uparrow$\hfil}%
                    \typeout{marginpar: $\uparrow$}\ignorespaces}
\def\mda{\marginpar[\boldmath\hfil$\downarrow$]%
                   {\boldmath$\downarrow$\hfil}%
                    \typeout{marginpar: $\downarrow$}\ignorespaces}
\def\mla{\marginpar[\boldmath\hfil$\rightarrow$]%
                   {\boldmath$\leftarrow $\hfil}%
                    \typeout{marginpar:
$\leftrightarrow$}\ignorespaces}
\def\Mua{\marginpar[\boldmath\hfil$\Uparrow$]%
                   {\boldmath$\Uparrow$\hfil}%
                    \typeout{marginpar: $\Uparrow$}\ignorespaces}
\def\Mda{\marginpar[\boldmath\hfil$\Downarrow$]%
                   {\boldmath$\Downarrow$\hfil}%
                    \typeout{marginpar: $\Downarrow$}\ignorespaces}
\def\Mla{\marginpar[\boldmath\hfil$\Rightarrow$]%
                   {\boldmath$\Leftarrow $\hfil}%
                    \typeout{marginpar:
$\Leftrightarrow$}\ignorespaces}
\overfullrule 5pt
\oddsidemargin -15mm
\marginparwidth 29mm
}

\begin{document}

\renewcommand{\bottomfraction}{0.1}
\renewcommand{\textfraction}{0}
\renewcommand{\floatpagefraction}{0.9}

\thispagestyle{empty}

\null
\hfill KA-TP-22-1998\\
\null
\hfill hep-ph/9901317\\
\vskip .8cm
\begin{center}
{\Large \bf Higher-order Results for Precision Observables\\[.5em]
in the Standard Model and the MSSM%
\footnote{Talk presented at the IVth International Symposium on Radiative
Corrections\\ (RADCOR~98), Barcelona, September 8--12, 1998.}
}
\vskip 2.5em
{\large
{\sc Georg Weiglein}\\[1ex]
{\normalsize \it Institut f\"ur Theoretische Physik, Universit\"at
Karlsruhe,\\
D-76128 Karlsruhe, Germany}
}
\vskip 2em
\end{center} \par
\vskip 1.6cm
\vfil
\begin{center}
{\bf Abstract} \\[.8em]
\end{center}
Recent higher-order results for precision observables in the Standard
Model (SM) and the Minimal Supersymmetric Standard Model (MSSM) 
and for the neutral $\cp$-even Higgs-boson masses of the MSSM are summarized. 
Furthermore a brief discussion of the technical aspects of 
evaluating higher-order corrections in the electroweak theory is given.
In the SM, results for the Higgs-mass dependence of the precision observables 
are analyzed. The exact two-loop results for the Higgs-mass dependence
of the fermionic contributions are compared with the results of an
expansion in the top-quark mass up to next-to-leading order.
In the MSSM, results for the leading two-loop contributions to the 
precision observables and to the masses of the neutral $\cp$-even Higgs
bosons are discussed. The latter are compared with results obtained by
renormalization group calculations.
\par
\vskip 1cm
\null
\setcounter{page}{0}
\clearpage

\title{HIGHER-ORDER RESULTS FOR PRECISION OBSERVABLES IN THE
STANDARD MODEL AND THE MSSM
}

\author{G.~WEIGLEIN}

\address{Institut f\"ur Theoretische Physik, Universit\"at Karlsruhe,\\
D--76128 Karlsruhe, Germany\\
E-mail: georg@particle.physik.uni-karlsruhe.de}


\maketitle\abstracts{
Recent higher-order results for precision observables in the Standard
Model (SM) and the Minimal Supersymmetric Standard Model (MSSM) 
and for the neutral $\cp$-even Higgs-boson masses of the MSSM are summarized. 
Furthermore a brief discussion of the technical aspects of 
evaluating higher-order corrections in the electroweak theory is given.
In the SM, results for the Higgs-mass dependence of the precision observables 
are analyzed. The exact two-loop results for the Higgs-mass dependence
of the fermionic contributions are compared with the results of an
expansion in the top-quark mass up to next-to-leading order.
In the MSSM, results for the leading two-loop contributions to the 
precision observables and to the masses of the neutral $\cp$-even Higgs
bosons are discussed. The latter are compared with results obtained by
renormalization group calculations.
}
 
\section{Introduction}

By comparing the electroweak Standard Model (SM) and its extensions,
most notably the Minimal Supersymmetric Standard Model (MSSM), with the
precision data~\cite{datasum98} it is possible to test the theory at
its quantum level, where all parameters of the model enter. In this way
one is able within the SM to infer indirect constraints on the mass of
the Higgs boson, which is the last missing ingredient of the SM. 
Within the MSSM the comparison with the precision data allows to
constrain mainly the parameters of the scalar quark sector.
Eventually the precision tests of the electroweak theory could hint
towards a distinction between the SM and Supersymmetry via their
respective virtual effects.
At present the bounds that can be obtained from this analysis on the 
parameters of the SM and the MSSM are still
relatively weak~\cite{datasum98}. In order to improve this situation, a
very high accuracy both of the measurements and the theoretical
predictions is needed. 

While QCD corrections to $\De r$ and the Z-boson observables 
$\sweff$ and $\Ga_l$ are known within the SM at
${\cal O}(\al \alps)$~\cite{qcd2} and ${\cal O}(\al \alps^2)$~\cite{qcd3}, 
only partial information is available about the electroweak two-loop 
contributions. Beyond one-loop order the resummations of the
leading one-loop contributions are known~\cite{resum,rheinsb}, the leading and
next-to-leading term in an expansion for asymptotically large values of
the top-quark mass, $\Mt$, have been evaluated~\cite{mt4,mt2,DGS}, and
also the leading term of an asymptotic expansion in the Higgs-boson mass,
$\MH$, has been derived~\cite{mh2}. 
The terms obtained in the $\Mt$-expansion were found to be numerically
sizable, and the next-to-leading term turned out to be about
equally large as the (formally) leading term~\cite{mt2}. Exact results 
have been derived for the Higgs-mass dependence of the fermionic 
two-loop corrections to the precision observables~\cite{mhdepDeR,rheinsb,bsw}.

In order to treat the MSSM at the same level of accuracy as the SM, 
the one-loop results for $\De r$ and the Z-boson observables~\cite{susyprec}
have to be supplemented by higher-order contributions. Recently the QCD
corrections to the $\rho$~parameter in the MSSM have been
evaluated~\cite{susydelrho}, which incorporates the leading two-loop
contribution to the precision observables. 

Besides the indirect tests of the MSSM via the precision observables,
there exists a very stringent direct test of the model, since it
predicts the existence of a light Higgs boson, which at the tree level
is restricted to be lighter than the Z~boson. A precise prediction for
the mass of the lightest Higgs boson, $\mh$, in terms of the relevant
SUSY parameters is crucial in order to determine the
discovery and exclusion potential of LEP2 and the upgraded Tevatron.
If the Higgs boson exists, it will be accessible at the LHC and future
linear colliders, where then a high-precision
measurement of the mass of this particle will become feasible.
A precise knowledge of the mass of the heavier $\cp$-even Higgs boson,
$\mH$, is important for
resolving the mass splitting between the $\cp$-even and -odd
Higgs-boson masses.

The tree-level bound on $\mh$ is strongly affected by the inclusion of
radiative corrections, which yield an upper bound of about $130$~GeV.
In order to go beyond the one-loop 
results~\cite{mhiggs1l,mhiggs1lfullb,mhiggs1lfull,mhiggs1lbmpz}, 
renormalization 
group (RG) methods have been applied for obtaining leading logarithmic 
higher-order contributions~\cite{mhiggsRGa,mhiggsRGb}. 
In the effective potential
approach the leading QCD corrections have been
calculated~\cite{mhiggsEffPot}. Up to now phenomenological
analyses have been based either on the results of the RG improved \onel\ 
effective potential approach~\cite{mhiggsRGb} or on the complete one-loop 
on-shell results~\cite{mhiggs1lfullb,mhiggs1lfull}, 
which differ by up to $20$~GeV in $\mh$. Recently a Feynman-diagrammatic 
calculation of the leading QCD corrections to the masses of the
neutral $\cp$-even Higgs bosons has been 
performed~\cite{mhiggs2la,mhiggs2lb}. The result obtained in this way
contains the leading two-loop corrections, the full
diagrammatic one-loop on-shell result~\cite{mhiggs1lfull}, and
further improvements taking into account 
leading electroweak two-loop and leading QCD corrections beyond $\oaas$.

In this paper some recent higher-order results in the SM and the MSSM are
summarized. Before discussing these results in some detail, a brief 
overview is given over technical issues involved in the calculation of 
two-loop corrections in the electroweak theory. 

\section{Techniques for calculating higher-order corrections in the
Standard Model and the MSSM}

The evaluation of higher-order corrections in the electroweak theory 
to the precision observables and to the predictions for the MSSM 
Higgs-boson masses involves as a main technical obstacle the evaluation
of two-loop 2-point functions. In order to express the results for the
precision observables in terms of the physical masses $\MW$ and $\MZ$ of the 
gauge bosons, the respective 2-point functions have to be evaluated
on-shell, i.e.\ at non-zero momentum transfer, while vertex and box
contributions in processes with light external particles can often be
reduced to vacuum integrals. As a consequence of the many different mass 
scales present in the electroweak theory, the evaluation of diagrams at the
two-loop level is in general very complicated. 

Besides the large number of contributing Feynman diagrams, problems 
encountered in such a calculation are due to the complicated tensor structure 
of the diagrams and to the fact that the scalar two-loop integrals are in
general not expressible in terms of polylogarithmic
functions~\cite{ScharfDipl}. Further complications are related to the
need for an adequate regularization and for a renormalization at the
two-loop level, which has not yet been worked out in full detail.

Concerning the regularization, two schemes are commonly applied in
calculations within the electroweak theory, namely Dimensional
Regularization (DREG)~\cite{dreg,ga5HV} and Dimensional
Reduction(DRED)~\cite{dred}. In DREG the regularization is performed by
analytically continuing the space-time dimension from 
4 to $D$. This prescription preserves the Lorentz and the gauge
invariance of the theory, apart from problems related to the treatment 
of $\ga_5$ in dimensions other than 4. In Supersymmetric theories,
however, a $D$-dimensional treatment of vector fields leads to a
mismatch between the fermionic and bosonic degrees of freedom, which
gives rise to a breaking of the Supersymmetric relations. This led to
the development of DRED. In this scheme only the momentum integrals are
treated $D$-dimensional, while the fields and the Dirac algebra are
kept 4-dimensional. DRED involves potential ambiguities related to the 
treatment of $\ga_5$, and its application to non-supersymmetric theories
turns out to be problematic (for a recent review see~\cite{jackjones}). 
In a naive application of both regularization schemes (without an
appropriate shift in the couplings which relates the two schemes to
each other) at one-loop order
differences in the finite parts of the Feynman diagrams (proportional
to $(D - 4)^0)$ arise, while at two-loop order both the finite parts and
the divergent part proportional to $(D - 4)^{-1}$ are different.

As already mentioned, a problem closely related to the one of finding
an adequate regularization for the considered calculation is the treatment of
$\ga_5$ in $D$ dimensions. The consistent prescription according
to~\cite{ga5HV,ga5BM} leads to a breaking of the Ward identities. As in all
cases where one works with a non-invariant regulator, this breaking has
to be compensated by the introduction of additional counterterms that
restore the symmetries. In order to investigate the invariance
properties of the regulators in Supersymmetry, a detailed study of the
supersymmetric Ward identities is necessary. A new approach that could
be helpful for applications in the electroweak theory is differential
renormalization~\cite{diffren}.

The results presented in this paper have been obtained using several
com\-puter-alge\-braic tools~\cite{sbaugw1}. The generation of the diagrams
and counterterm contributions has been done with the help of the
computer-algebra program \fea~\cite{fea}. For the calculations in the
MSSM, the relevant part of the MSSM Lagrangian has been implemented 
into \fea. The program inserts propagators and vertices
into the graphs in all possible ways and creates the amplitudes
including all symmetry factors. The calculation of the two-loop
diagrams and counterterms was performed with the package
\two~\cite{two}. It is based on a general algorithm for the tensor
reduction of two-loop 2-point functions, and reduces the amplitudes to
a minimal set of standard scalar integrals. Properties like the
gauge-parameter dependence or the validity of Ward identities can
directly be read off from the algebraic result obtained with \two.

For the calculations in the SM, DREG has been applied, while the results
within the MSSM presented below have been obtained using DRED. 
For the calculations described in this paper the use of an anticommuting 
$\ga_5$ in $D$ dimensions was possible without encountering inconsistencies.
The renormalization has been carried out using the on-shell
scheme. Within the SM a two-loop renormalization was necessary 
in the gauge-boson sector. In
the MSSM a two-loop renormalization in the Higgs-boson sector and a 
one-loop renormalization of the gauge-boson and scalar quark sector had
to be performed.
Those two-loop scalar integrals for which no analytic expression
in terms of polylogarithmic functions can be derived have been
evaluated numerically with one-dimensional integral
representations~\cite{intnum}. These allow a very fast calculation of
the integrals with high precision without any approximation in the
masses.

\section{Higgs-mass dependence of precision observables in the Standard
Model at two-loop order}

In order to study the Higgs-mass dependence of the precision
observables $\De r$, $\sweff$ and $\Ga_l$ at two-loop order,
we consider subtracted quantities of the form
\beq
a_{\mathrm{subtr}}(\MH) = a(\MH) - a(M^0_{\PH}), \quad \mbox{where }
a = \De r, \sweff, \Ga_l,
\eeq
which indicate the shift in the precision observables caused by varying
the Higgs-boson mass between $M^0_{\PH}$ and $\MH$. 
In the analysis below the Higgs-boson mass is varied in the interval 
$65 \, \, \mbox{GeV} \leq \MH \leq 1 \TeV$. 
The quantity $\De r$, which determines the relation between the
vector-boson masses in terms of the Fermi constant, is derived from 
muon decay~\cite{sirlin}. It is defined according to
\beq
\MW^2 \left(1 - \frac{\MW^2}{\MZ^2}\right) =
\frac{\pi \al}{\sqrt{2} \GF} \left(1 + \De r\right).
\eeq
The leptonic effective weak mixing angle and the leptonic width of the
Z~boson are determined from the effective couplings of the neutral
current at the Z-boson resonance, 
$J_{\mu}^{\mathrm{NC}} = \left( \sqrt{2} \GF \MZ^2 \right)^{1/2}
\left[g_V^f \ga_{\mu} - g_A^f \ga_{\mu} \ga_5 \right]$, according to
\beq
\sweff = \frac{1}{4 |Q_f|} \left(1 - \frac{\Re g_V^f}{\Re g_A^f}
\right) , \quad
\Ga_l = \frac{\al \MZ}{12 \sw^2 \cw^2}
\left(1 + \frac{3 \al}{4 \pi}\right) \left(|g_V^f|^2 + |g_A^f|^2\right) .
\eeq

Potentially large $\MH$-dependent contributions are the ones associated
with the top quark due to the large Yukawa coupling $\Pt \bar{\Pt} \PH$,
and the corrections proportional to $\De\al$. Further contributions are
the ones of the light fermions (except the corrections already contained
in $\De\al$), and purely bosonic contributions. The latter are expected
to be relatively small~\cite{mhdepDeR}. We therefore have focussed on the
Higgs-dependent fermionic contributions to the precision observables,
for which we have obtained exact two-loop results~\cite{mhdepDeR,bsw}.

\begin{figure}
\begin{center}
\epsfig{figure=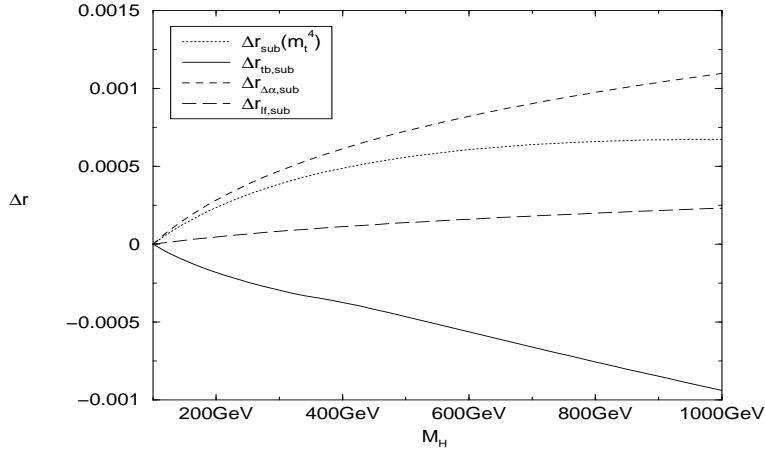, width=10cm, height=6cm}
\end{center}
\caption{\label{fig:drmh2l}
Higgs-mass dependent fermionic contributions to $\De r$ at two-loop
order. The different curves show the contribution from the diagrams
involving the top/bottom doublet ($\De r_{\mathrm{tb}}$), the
contribution
proportional to $\De\al$ ($\De r_{\De\al}$), the contribution of the
light fermions ($\De r_{\mathrm{lf}}$), and the approximation of the
top/bottom correction by the leading term proportional to $\Mt^4$
($\De r(\Mt^4)$).
}
\end{figure}

\begin{figure}
\begin{center}
\epsfig{figure=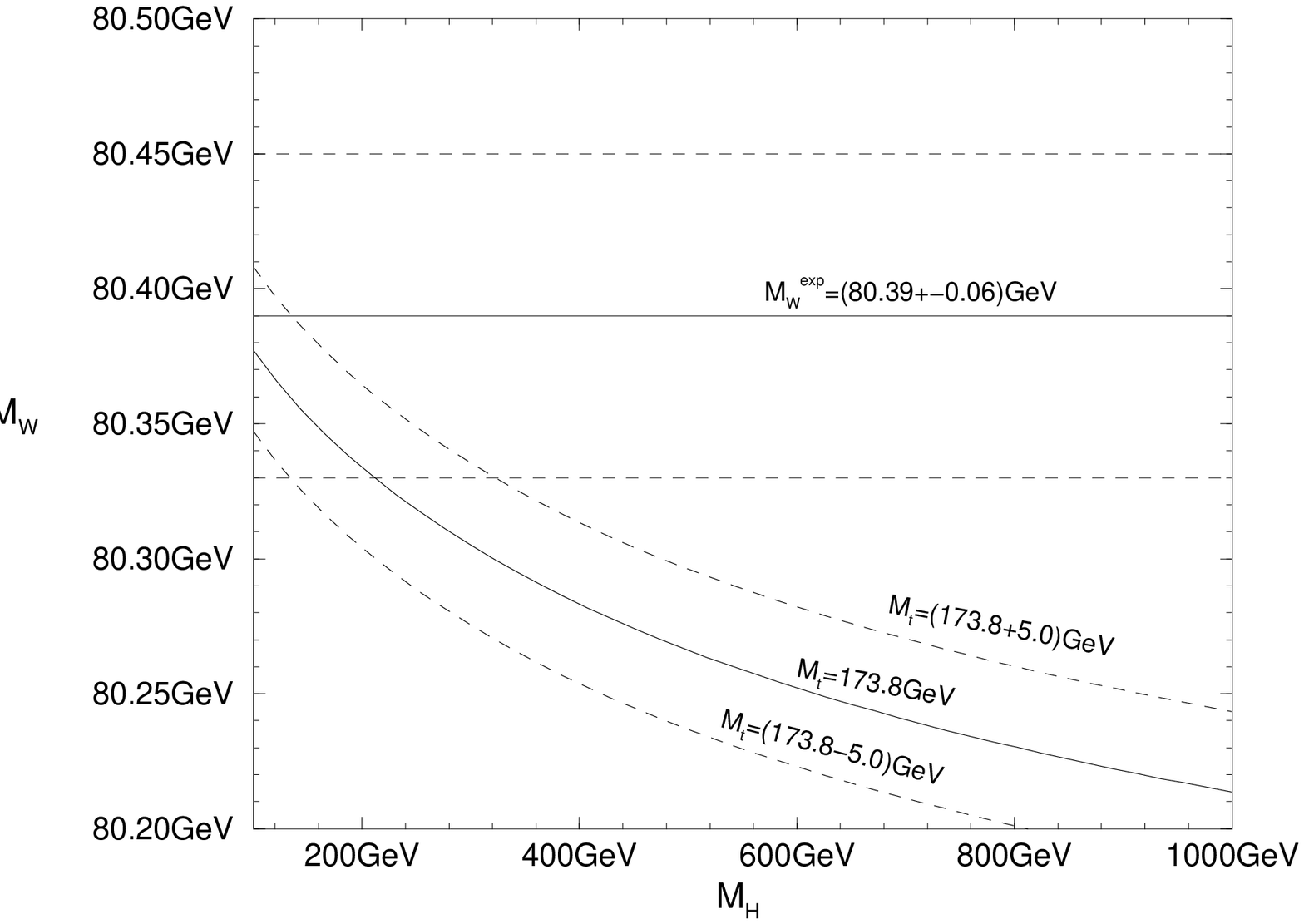, width=9cm, height=5.3cm}\\
\epsfig{figure=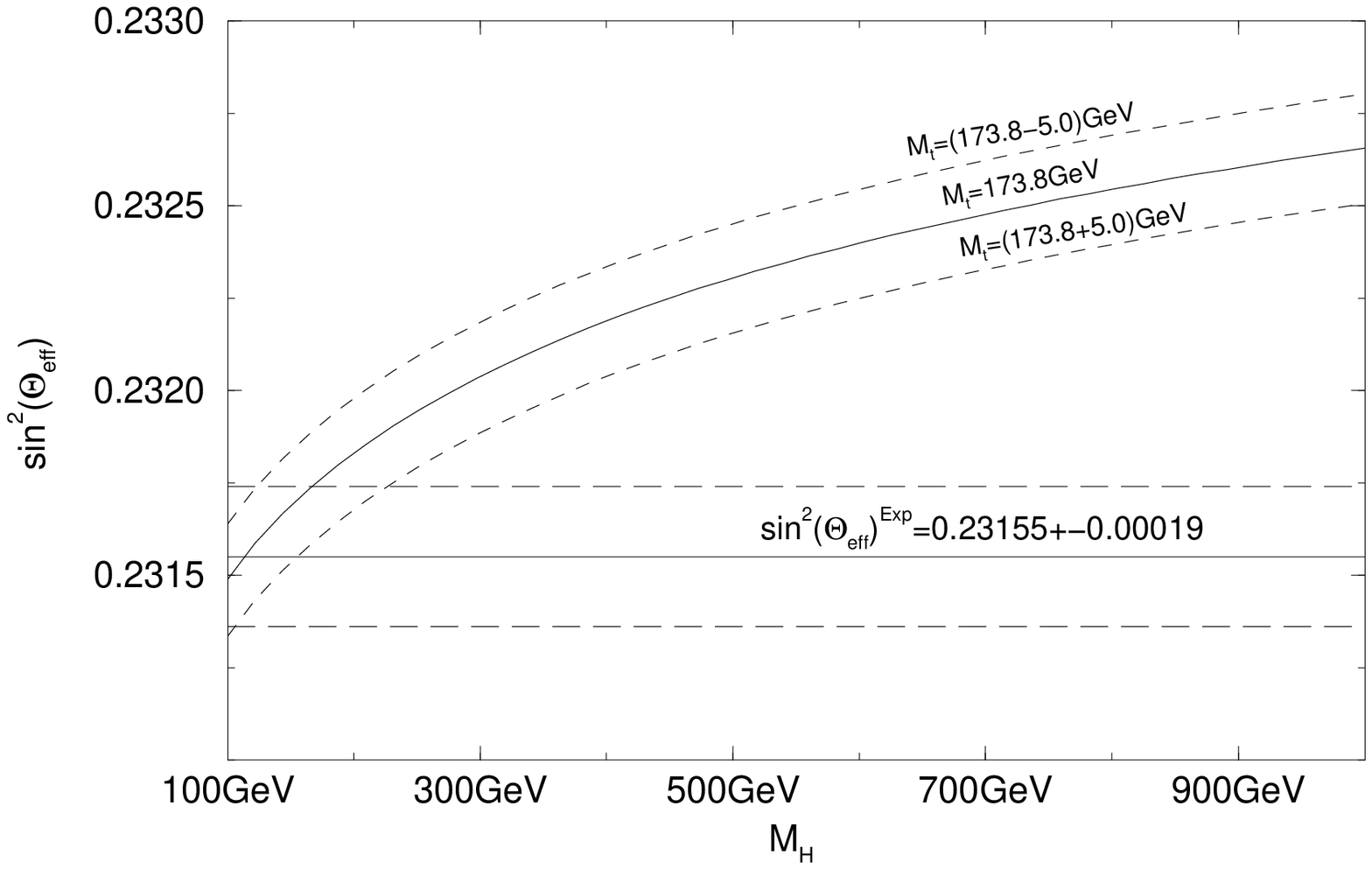, width=9cm, height=5.3cm}\\
\epsfig{figure=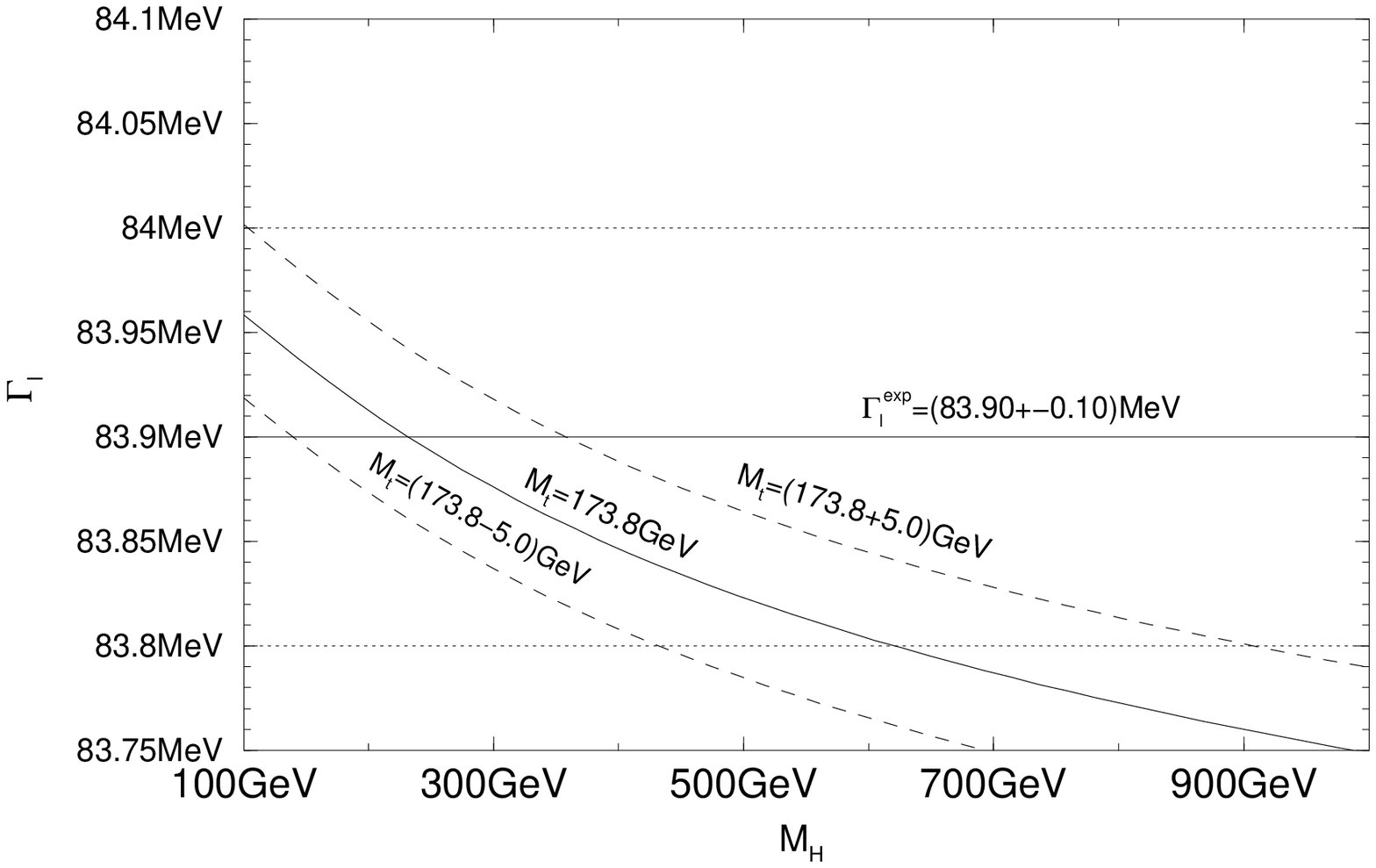, width=9cm, height=5.3cm}\\[-1em]
\end{center}
\caption{\label{fig:obsmh}
The Higgs-mass dependence of the prediction for
$\MW$, $\sweff$ and $\Ga_l$ is shown for
different values of $\Mt$. The experimental values of the observables
are also indicated.
}
\end{figure}

\reffi{fig:drmh2l} shows the Higgs-mass dependence of the two-loop
corrections to $\De r$ associated with the top/bottom doublet, with
$\De\al$, and with the light fermions. The dotted line furthermore
indicates the Higgs-mass dependence of the leading
$\Mt^4$-term~\cite{mt4} in the top/bottom contribution. 
The two-loop top/bottom contribution gives rise to a shift in the
W-boson mass of 
$\De M^{\mathrm{top}}_{\PW, \mathrm{subtr}, (2)}(\MH = 1000 \GeV) \approx 
16$~MeV, which is about $10\%$ of the one-loop contribution.
The Higgs-mass dependence of 
this contribution turns
out to be very poorly approximated by the leading $\Mt^4$-term; the
contribution of the latter even enters with a different sign. It can be seen
from \reffi{fig:drmh2l} that the two-loop contributions to a large
extent cancel each other. The contribution of the light fermions yields
a shift in $\MW$ of up to 
4~MeV. In total the
two-loop contributions lead to a slight increase in the sensitivity of
$\De r$ to the Higgs-boson mass compared to the one-loop case.
Similarly as for $\De r$, also for the Higgs-mass
dependence of $\sweff$ and $\Ga_l$ large cancellations occur between the
two-loop contributions. For these observables the higher-order
contributions decrease the sensitivity to the Higgs-boson
mass~\cite{bsw}. The Higgs-mass dependence of $\MW$, $\sweff$ and
$\Ga_l$ is shown in \reffi{fig:obsmh} together with the experimental 
values of the observables. The theoretical values for $\MH^0 = 65$~GeV are
taken from~\cite{DGS}.

In \refta{tab:Obscomp} the Higgs-mass
dependence of $\MW$ and $\sweff$ based on the exact result for 
the fermionic contributions is compared with the results of the
expansion in the top-quark mass up to ${\cal O}(\GF^2 \Mt^2 \MZ^2)$
given in \cite{DGS} (with the input parameters as in \cite{DGS}).%
\footnote{It should be noted that in contrast to the numbers given
in~\cite{rheinsb} the contribution of the light fermions is included here.}
Over the range of Higgs-mass values from 65~GeV to 1~TeV the difference
between the results amounts to about 4~MeV for $\MW$ and to about 
$7 \cdot 10^{-5}$ for $\sweff$. The difference in the prediction for
$\sweff$ turns out to be mainly induced by the difference in the
prediction for $\MW(\MH)$. Evaluating
$\sweffsub(\MH = 1 \TeV)$ using the value for
$\MW(\MH = 1 \TeV)$ from \cite{DGS} instead of our result (see
\refta{tab:Obscomp})
yields a value for $\sweffsub(\MH = 1 \TeV)$ that
differs from the one given in \cite{DGS} only by about 
$2 \cdot 10^{-5}$.

\begin{table}
\caption{\label{tab:Obscomp}
The Higgs-mass dependence of $\MW$ and $\sweff$
based on the exact result for the
fermionic contribution (left column) and on the result
of the expansion in $\Mt$ (right column).
}
$$
\begin{array}{|c||c|c||c|} \hline
\MH /\GeV &                
\MWsub^{\mathrm{ferm}} /\MeV &
\MWsub^{\mathrm{top}, \De\al, \mathrm{DGS}} /\MeV &
\De \MW /\MeV\\ \hline
65   & 0    & 0    & 0   \\ \hline
100  & - 23  & - 23  & 0 \\ \hline
300  & - 95  & - 98  & 3 \\ \hline
600  & - 148 & - 152 & 4 \\ \hline
1000 & - 187 & - 191 & 4 \\ \hline
\multicolumn{4}{c}{}\\ \hline
\MH /\GeV &
\sweffsub^{\mathrm{ferm}} &
\sweffsub^{\mathrm{top}, \De\al, \mathrm{DGS}} &
\De \sweff \\ \hline
65   & 0    & 0    & 0   \\ \hline
100  & 0.00020  & 0.00021  & 0.00001 \\ \hline
300  & 0.00076  & 0.00080  & 0.00004 \\ \hline
600  & 0.00112  & 0.00118  & 0.00006 \\ \hline
1000 & 0.00138  & 0.00145  & 0.00007 \\ \hline
\end{array}
$$
\end{table}


In the comparison of \refta{tab:Obscomp} besides
the difference between the exact treatment of the top-quark
contributions and the expansion up to next-to-leading order in $\Mt$
further effects enter. In~\cite{DGS} the Higgs-mass dependence of 
light-fermion contributions only arises from reducible contributions that are
generated by Dyson summation, while in our results above also the 
Higgs-mass dependence of the light-fermion contributions is taken into
account exactly. Further differences are due to a different treatment of
higher-order  QCD and electroweak corrections. In~\cite{gsw} the effect solely
caused by the difference between the exact treatment of the top-quark
contributions and the expansion in $\Mt$ has been analyzed in detail,
and it has been found that over the full range of Higgs-boson masses
this effect alone amounts to a difference of about 2~MeV in $\MW$ and
$3 \cdot 10^{-5}$ in $\sweff$. In the case of $\MW$ the three other
sources of differences mentioned above, namely the light-fermion
contribution and the treatment of higher-order QCD and electroweak
corrections, all individually give rise to a difference of $2-3$~MeV
over the full range of Higgs-boson masses, which combine with the 
effect of the $\Mt$-expansion to the difference of 4~MeV listed in the
last entry of the first table in \refta{tab:Obscomp}.

\section{QCD corrections to precision observables in the
MSSM}

The leading two-loop corrections to the electroweak precision 
observables in the
MSSM enter via the quantity $\De\rho$, which is given in terms of the
transverse parts of the two-loop Z-boson and W-boson self-energies as
\beq
\Delta \rho =
\frac{\Sigma^{\PZ}(0)}{\MZ^2} - \frac{\Sigma^{\PW}(0)}{\MW^2} .
\eeq
The contributions to $\De\rho$ of $\oaas$ have been evaluated 
in~\cite{susydelrho}. They can be separated into the contribution of
diagrams with gluon exchange, which dominates in general (it is
of the order of 10--15\% of the one-loop result) and for which a 
very compact analytic result can be derived~\cite{susydelrho}, and into
the gluino-exchange contribution, which goes to zero for large gluino
mass. 

\begin{figure}[ht!]
\begin{center}
\mbox{
\psfig{figure=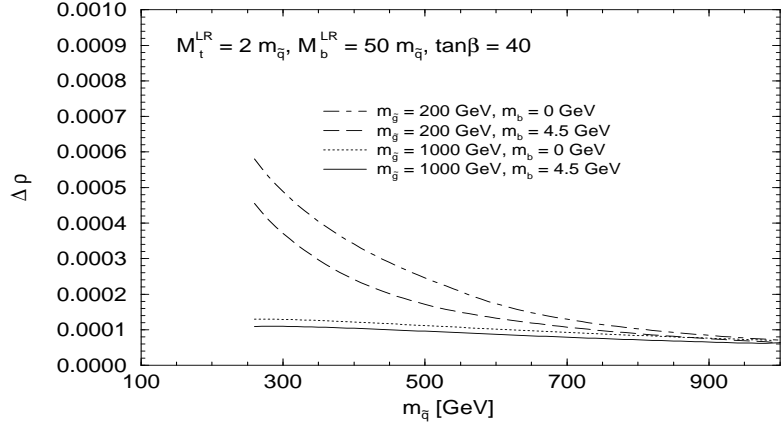,width=6.3cm,height=5.7cm,
                      bbllx=150pt,bblly=100pt,bburx=450pt,bbury=420pt}}
\end{center}
\caption[]{
Contribution of the gluino-exchange diagrams to $\De\rho$ for $\Tb =
40$ and large mixing in the scalar bottom sector. The full result is
compared to the case where mixing in the scalar bottom sector is
neglected ($\Mb = 0$).
}
\label{fig:drhogluino}
\end{figure}

The results of~\cite{susydelrho} have recently been extended by 
including also the effects of mixing in the scalar bottom sector, which
previously had been neglected, and by evaluating the gluon-exchange
contributions to the precision observables $\De r$, $\sweff$ and
$\Ga_l$, thus going beyond the $\De\rho$~approximation~\cite{hhwprec}.
In \reffi{fig:drhogluino} the gluino-exchange contribution is shown 
for $\Tb = 40$ and large mixing in the scalar bottom sector as
a function of the common scalar mass parameter 
\beq
m_{\tilde q} \equiv
M_{\tilde{t}_{L}}=M_{\tilde{t}_{R}}=M_{\tilde{b}_{L}} =
M_{\tilde{b}_{R}}.
\label{eq:susyrel}
\eeq
The $M_{\tilde{q}_{L/R}}$ are the soft SUSY
breaking parameters in the diagonal entries
of the mass matrices of the scalar top and bottom quarks. The
off-diagonal entries read $m_q M_q^{LR}$ with $M^{LR}_t = A_t - \mu \,
\CTb$, $M^{LR}_b = A_b - \mu \Tb$, see~\cite{susydelrho}. 
The full result is compared in \reffi{fig:drhogluino} 
to the case where mixing in
the scalar bottom sector is neglected ($\Mb = 0$). The figure shows
that effects of $\Sbot$~mixing can be sizable for large $\Tb$ and 
small gluino masses.

\begin{figure}[ht!]
\begin{center}
\mbox{
\psfig{figure=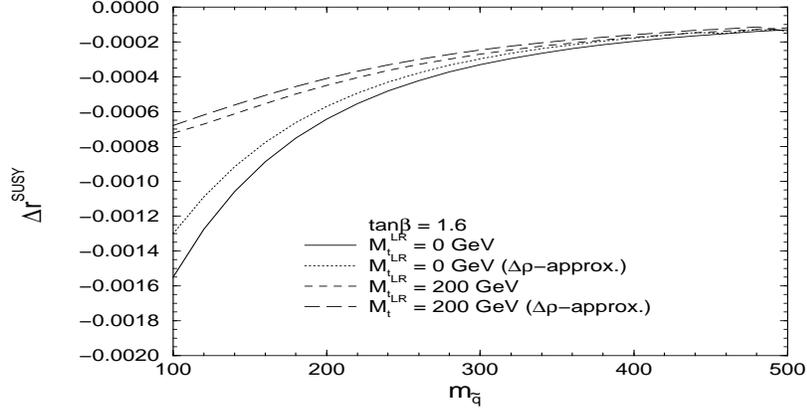,width=10.6cm,height=5.5cm,
              bbllx=25pt,bblly=85pt,bburx=550pt,bbury=445pt}}
\parbox{12cm}{
\caption[]{Contribution of the gluon-exchange diagrams to
$\Delta r^{\rm SUSY}$ as a function of the common scalar mass parameter
$m_{\tilde q}$ for the scenario of no mixing
($M^{LR}_t=0$) and large mixing ($M_t^{LR}=200$~GeV)
in the $\Stop$ sector. The exact result is compared with the approximation
derived from the contribution of $\De\rho$.
\label{fig:deltrglu}
}}
\end{center}
\end{figure}

The leading contribution to $\De r$ in the MSSM can be approximated by
the contribution to $\De\rho$ according to $\De r \approx - \cw^2/\sw^2
\De \rho$, where $\cw^2 = 1 - \sw^2 = \MW^2/\MZ^2$. The gluon-exchange 
contribution to $\De r^{\rm SUSY}$ of a squark doublet is given by
\beq
\Delta r^{\rm SUSY}_{\rm gluon} = \Pi^{\gamma}(0) -
\frac{\cw^2}{\sw^2} \left(\frac{\de \MZ^2}{\MZ^2} -
\frac{\de \MW^2}{\MW^2} \right) +
\frac{\Sigma^\PW(0) - \de \MW^2}{\MW^2},
\label{eq:deltrglu}
\eeq
where $\de \MW^2 = \Re \Sigma^{\PW}(\MW^2)$,
$\de \MZ^2 = \Re \Sigma^{\PZ}(\MZ^2)$, and $\Pi^{\gamma}$,
$\Sigma^{\PW}$, and $\Sigma^{\PZ}$ denote the transverse parts of
the two-loop gluon-exchange contributions to the photon vacuum
polarization
and the W-boson and Z-boson \ses, respectively,
which all are understood to contain the subloop renormalization.
This result is shown together with the $\De\rho$~approximation in
\reffi{fig:deltrglu}. 
The two-loop contribution leads to a shift in the W-boson mass of up to
$20$~MeV for low values of $m_{\tilde q}$ in the no-mixing case. If the
parameter $M^{LR}_t$ is made large or the relation \refeq{eq:susyrel}
is relaxed, much bigger effects are possible~\cite{susydelrho}. As can
be seen in \reffi{fig:deltrglu}, the $\De\rho$ contribution
approximates the full result rather well. The two results agree within
10--15\%.

\section{Diagrammatic two-loop results for the masses of the neutral
$\cp$-even Higgs bosons in the MSSM}

In~\cite{mhiggs2la,mhiggs2lb} a Feynman-diagrammatic calculation 
of the leading QCD 
corrections to the masses of the neutral $\cp$-even Higgs bosons has been 
performed. The result obtained in this way contains the leading two-loop 
corrections, the complete diagrammatic one-loop on-shell 
result~\cite{mhiggs1lfull}, and further improvements taking into account 
leading electroweak two-loop and leading QCD corrections beyond $\oaas$. The
calculation of $\mh$ and $\mH$ has been carried out for arbitrary values of
the parameters in the Higgs and scalar top sector of the MSSM. The
results have been implemented into the Fortran program
\fh~\cite{FeynHiggs}.

\begin{figure}
\begin{center}
\mbox{
\psfig{figure=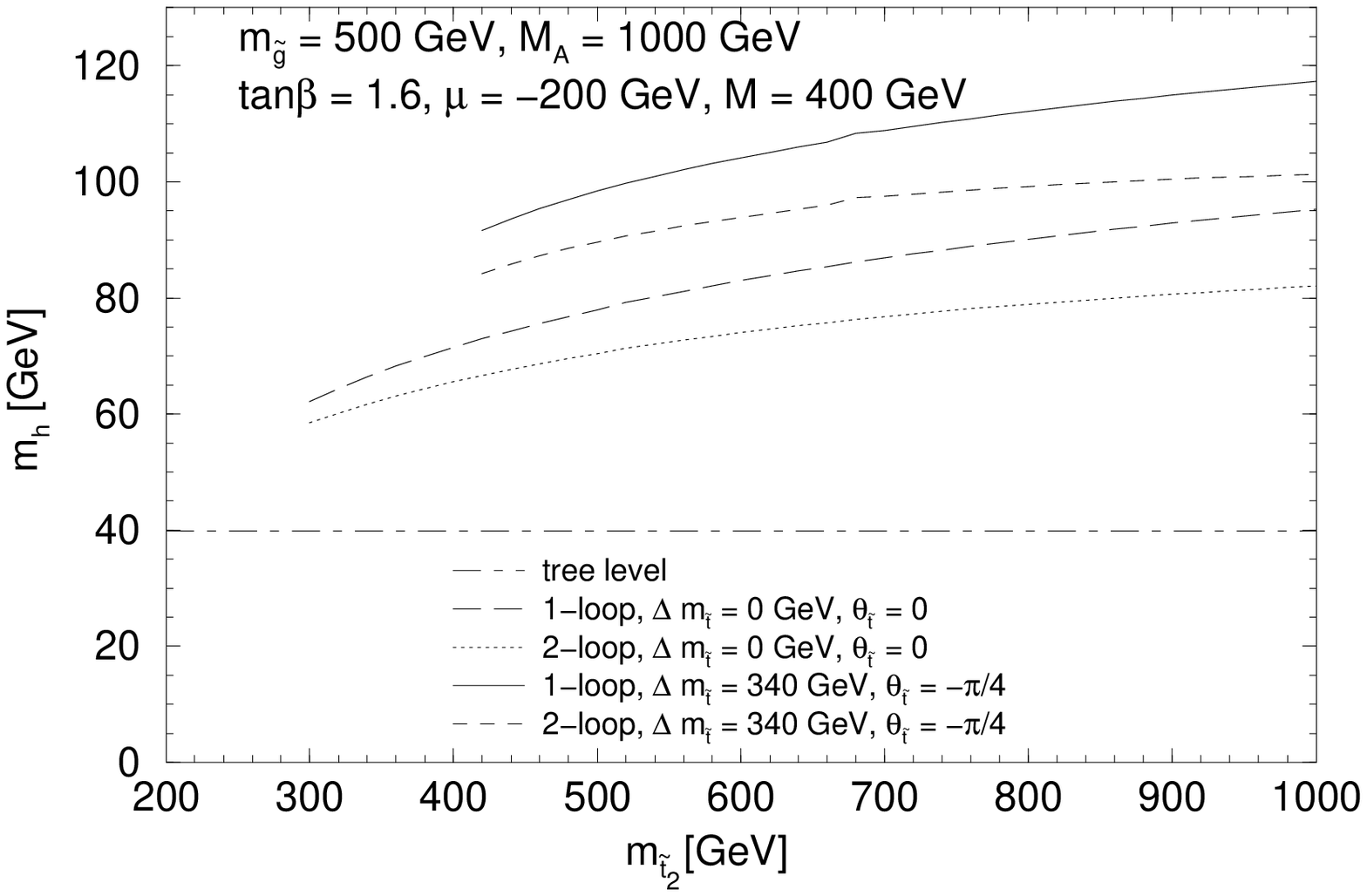,width=6.3cm,height=5.7cm,
                      bbllx=150pt,bblly=100pt,bburx=450pt,bbury=420pt}}
\end{center}
\caption[]{
The mass of the lightest Higgs boson in the MSSM in terms of the physical
parameters $\mste, \mstz$ and $\tst$, where
$\De\mst \equiv \mstz - \mste$. The scenarios
$\De\mst = 0$~GeV, $\tst = 0$ (no mixing) and
$\De\mst = 340$~GeV, $\tst = -\pi/4$ (maximal mixing) are shown.
}
\label{fig:mh_mst2}
\end{figure}

In \reffi{fig:mh_mst2} the mass of the lightest Higgs boson is shown as
a function of the mass of the heavier scalar top quark, $\mstz$, for
different values of $\mste$ and the mixing angle in the scalar top
sector, $\tst$. The tree-level, the one-loop and the 
two-loop results for $\mh$ are given
for the case of no mixing ($\tst = 0$) and degenerate $\Stop$ masses
($\De\mst \equiv \mstz - \mste = 0$), as well as for maximal mixing 
($\tst = -\pi/4$) and a mass difference $\De\mst \approx 340$~GeV that
gives rise to the largest values of $\mh$ (the corresponding
values for $\De\rho^{\mathrm{SUSY}}$ lie within the experimentally 
favored~\cite{delrhoexp} 
region $\De\rho^{\mathrm{SUSY}} \lsim 1.3 \cdot 10^{-3}$). 
As can be seen from the figure, in both cases
the two-loop corrections give rise to a large reduction of the one-loop
on-shell result of up to 15~GeV.

\begin{figure}
\begin{center}
\mbox{
\psfig{figure=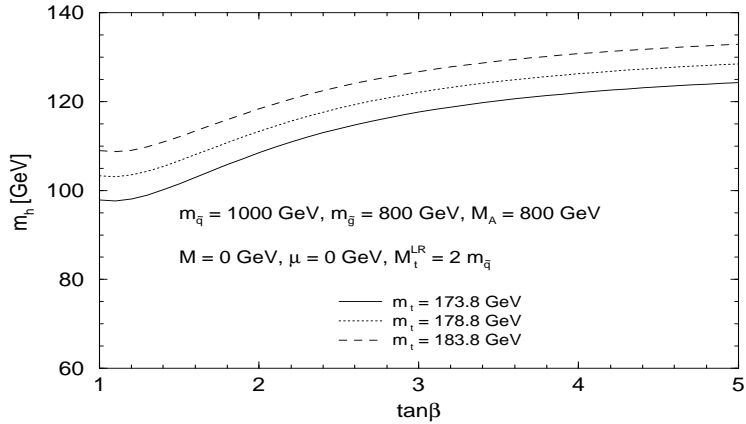,width=6.3cm,height=5.7cm,
                      bbllx=150pt,bblly=100pt,bburx=450pt,bbury=420pt}}
\end{center}
\caption[]{
The maximally possible value for $\mh$ as a function of $\Tb$ for 
$\msq = 1000$~GeV and three values of $\Mt$.
}
\label{fig:mhmax_tb_UP}
\end{figure}

In order to derive an upper bound for $\mh$ within the MSSM as a function of
$\tan\be$, we have performed a scan over the other parameters such that the
corresponding value of $\mh$ is maximized. \reffi{fig:mhmax_tb_UP}
shows the maximally possible value for $\mh$ in the region $\tan\be \leq 5$
for $\msq = 1000$~GeV and 
three values of the top-quark mass,
$\Mt = 173.8 \gev, 178.8 \gev, 183.8 \gev$ (the
current experimental value of $\Mt$ and values
one and two standard deviations above it). It can be seen that the 
maximal values of $\mh$ in the particularly
interesting region of $\tan\be \lsim 2$ are
at the edge of the LEP2 reach, which is expected to be roughly 105~GeV.
The question whether a full coverage of this region is possible depends
sensitively on the range of $\Mt$ values employed for evaluating~$\mh$.

\begin{figure}                                                         
\begin{center}
\mbox{
\psfig{figure=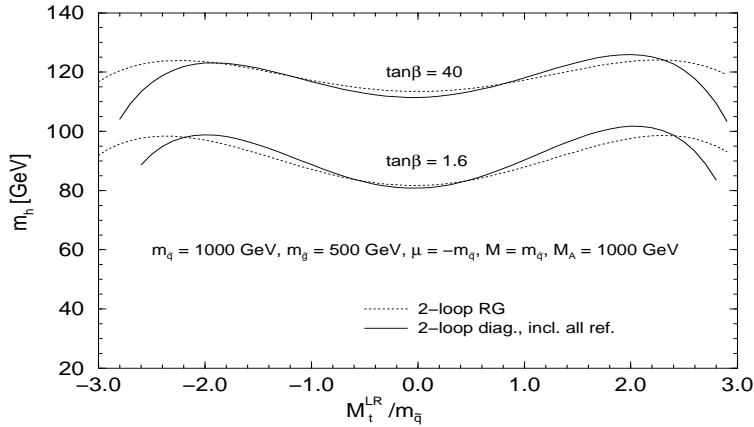,width=6.3cm,height=5.7cm,
                      bbllx=150pt,bblly=100pt,bburx=450pt,bbury=420pt}}
\end{center}
\caption[]{
Comparison between the Feynman-diagrammatic calculation and the results
obtained by renormalization group methods. The mass of the lightest
Higgs boson is shown for the two scenarios with
$\Tb = 1.6$ and $\Tb = 40$ as a function of $\Mtlr/\msq$.
}
\label{fig:mh_MtLRdivmq_MA1000_RGVergleich}
\end{figure}

In \reffi{fig:mh_MtLRdivmq_MA1000_RGVergleich} the results of the
Feynman-diagrammatic calculation for $\mh$ are compared with the
results obtained by RG methods~\cite{mhiggsRGb}. The mass of the
lightest Higgs boson is shown as a function of $\Mtlr/\msq$ for the 
two scenarios with $\Tb = 1.6$ and $\Tb = 40$. Good agreement is found
for vanishing mixing in the scalar top sector ($\Mtlr = 0$). Sizable
deviations occur, however, when mixing in the $\Stop$ sector is taken
into account. They reach about 5~GeV for moderate mixing and become
very large for $|\Mtlr/\msq| \gsim 2.5$. The maximal value for $\mh$ in
the diagrammatic approach is reached for $\Mtlr/\msq \approx \pm 2$,
whereas the RG results have a maximum at $\Mtlr/\msq \approx \pm 2.4$,
which corresponds to the maximum in the one-loop result. In the case of
positive $\Mtlr$, the maximal values for $\mh$ reached in the
diagrammatic calculation are up to $5~(3)\gev$ larger than the ones of
the RG method for $\Tb = 1.6~(40)$. A comparison of the diagrammatic
and the RG results in terms of the physical parameters $\mste$,
$\mstz$, $\tst$ has been performed in~\cite{mhiggs2lb}. It yields
qualitatively the same results.
An additional source of deviation between the Feynman-diagrammatic and the
RG result is the dependence of the diagrammatic result on the mass of
the gluino, which does not appear in the RG result. Its variation gives
rise to a shift of the diagrammatic result relative to the RG result 
of up to $\pm 2$~GeV.

\section*{Acknowledgements}
The author thanks S.~Bauberger, S.~Heinemeyer, W.~Hollik, and A.~Stremplat 
for collaboration on various
parts of the results presented here. I also want to thank J.~Sol\'a and
the Organizing Committee for the invitation and their kind hospitality
during RADCOR 98.


\section*{References}
\begin{thebibliography}{99}
\newcommand{\anp}[3]{{\sl Ann.~Phys.} {\bf #1} (19#2) #3}
\newcommand{\app}[3]{{\sl Acta~Phys.~Pol.} {\bf #1} (19#2) #3}
\newcommand{\cmp}[3]{{\sl Commun. Math. Phys.} {\bf #1} (19#2) #3}
\newcommand{\cpc}[3]{{\sl Comp. Phys. Commun.} {\bf #1} (19#2) #3}
\newcommand{\fp}[3]{{\sl Fortschr. Phys.} {\bf #1} (19#2) #3}
\newcommand{\ijmp}[3]{{\sl Int. J. Mod. Phys.} {\bf #1} (19#2) #3}
\newcommand{\jetp}[3]{{\sl JETP} {\bf #1} (19#2) #3}
\newcommand{\jetpl}[3]{{\sl JETP Lett.} {\bf #1} (19#2) #3}
\newcommand{\jmp}[3]{{\sl J. Math. Phys.} {\bf #1} (19#2) #3}
\newcommand{\mpl}[3]{{\sl Mod. Phys. Lett.} {\bf #1} (19#2) #3}
\newcommand{\nc}[3]{{\sl Nuovo Cimento} {\bf #1} (19#2) #3}
\newcommand{\ncl}[3]{{\sl Nuovo Cimento Lett.} {\bf #1} (19#2) #3}
\newcommand{\nim}[3]{{\sl Nucl. Instr. Meth.} {\bf #1} (19#2) #3}
\newcommand{\np}[3]{{\sl Nucl. Phys.} {\bf #1} (19#2)~#3}
\newcommand{\npB}[3]{{\sl Nucl. Phys.} {\bf B #1} (19#2)~#3}
\newcommand{\nphbps}[3]{{\sl Nucl. Phys.} {\bf B} {\it (Proc. Suppl.)}
{\bf #1B} (19#2) #3}
\newcommand{\plB}[3]{{\sl Phys. Lett.} {\bf B #1} (19#2) #3}
\newcommand{\prD}[3]{{\sl Phys. Rev.} {\bf D #1} (19#2) #3}
\newcommand{\prl}[3]{{\sl Phys. Rev. Lett.} {\bf #1} (19#2) #3}
\newcommand{\pl}[3]{{\sl Phys. Lett.} {\bf #1} (19#2) #3}
\newcommand{\ptp}[3]{{\sl Prog. Theo. Phys.} {\bf #1} (19#2) #3}
\newcommand{\sptp}[3]{{\sl Suppl. Prog. Theo. Phys.} {\bf #1} (19#2) #3}
\newcommand{\sjnp}[3]{{\sl Sov. J. Nucl. Phys.} {\bf #1} (19#2) #3}
\newcommand{\zp}[3]{{\sl Z. Phys.} {\bf #1} (19#2) #3}
\newcommand{\vj}[4]{{\sl #1~}{\bf #2} (19#3) #4}
\newcommand{\ej}[3]{{\bf #1} (19#2) #3}
\newcommand{\vjs}[2]{{\sl #1~}{\bf #2}}

\bibitem{datasum98}
F.~Teubert, these proceedings, hep-ph/9811414.

\bibitem{qcd2}
A.~Djouadi and C.~Verzegnassi, \pl{B 195}{87}{265};\\
A.~Djouadi, \nc{A 100}{88}{357};\\
B.A.~Kniehl, \np{B 347}{90}{89};\\
F.\ Halzen and B.A.~Kniehl, \npB{353}{91}{567}.

\bibitem{qcd3}
L.~Avdeev, J.~Fleischer, S.M.~Mikhailov and O.~Tarasov,
\plB{336}{94}{560}; E: \plB{349}{95}{597};\\
K.~Chetyrkin, J.~K\"uhn and M.~Steinhauser, \plB{351}{95}{331};
\prl{75}{95}{3394}.

\bibitem{resum}
W.\ Marciano, \prD{20}{79}{274};\\
A.\ Sirlin, \prD{29}{84}{89};\\
M.\ Consoli, W.\ Hollik, F.\ Jegerlehner, \plB{227}{89}{167}.

\bibitem{rheinsb}
G.~Weiglein, \app{B 29}{98}{2735}.

\bibitem{mt4}
 J.~van der Bij and F.~Hoogeveen, \npB{283}{87}{477};\\
R.~Barbieri, M.~Beccaria, P.~Ciafaloni, G.~Curci, A.~Vicere,
\plB{288}{92}{95}, E: {\bf B 312} (1993) 511;
\npB{409}{93}{105};\\
J.~Fleischer, O.~Tarasov and F.~Jegerlehner,
\plB{319}{93}{249}; \prD{51}{95}{3820}.

\bibitem{mt2}
G.~Degrassi, P.~Gambino and A.~Vicini, \plB{383}{96}{219}.

\bibitem{DGS}
G.\ Degrassi, P.\ Gambino and A.\ Sirlin, \plB{394}{97}{188};\\
G.\ Degrassi, P.\ Gambino, M.~Passera and A.~Sirlin, \plB{418}{98}{209}.

\bibitem{mh2}
J.~van der Bij and M.~Veltman, \npB{231}{84}{205}.

\bibitem{mhdepDeR}
S.~Bauberger and G.~Weiglein, \plB{419}{98}{333}.

\bibitem{bsw}
S.~Bauberger, A.~Stremplat and G.~Weiglein, in preparation.

\bibitem{susyprec}
P.~Chankowski, A.~Dabelstein, W.~Hollik, W.~M\"osle, S.~Pokorski
and J.~Rosiek, \np{B 417}{94}{101};\\
D.~Garcia and J.~Sol\`a, \mpl{A 9}{94}{211};\\
D.~Garcia, R.~Jim\'enez and J.~Sol\`a, {\sl Phys. Lett.} {\bf B 347}
(1995) 309; {\bf B 347} (1995) 321;\\
D.~Garcia and J.~Sol\`a, {\sl Phys. Lett.} {\bf B 357} (1995) 349;\\
A.~Dabelstein, W.~Hollik, W.~M\"osle, in {\em Perspectives for
Electroweak Interactions in $e^+e^-$ Collisions}, ed.\ B.~Kniehl
(World Sci., 1995), p.~345;\\
P.~Chankowski and S.~Pokorski, {\sl Nucl. Phys.} {\bf B 475} (1996) 3.

\bibitem{susydelrho}
A.~Djouadi, P.~Gambino, S.~Heinemeyer, W.~Hollik, C.~J\"unger and
G.~Weiglein, \prl{78}{97}{3626}; \prD{57}{98}{4179}.

\bibitem{mhiggs1l}
H.~Haber and R.~Hempfling, \prl{66}{91}{1815};\\
Y.~Okada, M.~Yamaguchi, T.~Yanagida, \ptp{85}{91}{1};\\
J.~Ellis, G.~Ridolfi and F.~Zwirner, \plB{257}{91}{83},
                                     \plB{262}{91}{477};\\
R.~Barbieri and M.~Frigeni, \plB{258}{91}{395}.

\bibitem{mhiggs1lfullb}
P.~Chankowski, S.~Pokorski and J.~Rosiek, \npB{423}{94}{437}.

\bibitem{mhiggs1lfull}
A.~Dabelstein, \npB{456}{95}{25}, \zp{C 67}{95}{495}.

\bibitem{mhiggs1lbmpz}
J.~Bagger, K.~Matchev, D.~Pierce, R.~Zhang, \npB{491}{97}{3}.

\bibitem{mhiggsRGa}
J.~Casas, J.~Espinosa, M.~Quir\'os, A.~Riotto, \npB{436}{95}{3}.

\bibitem{mhiggsRGb}
M.~Carena, J.~Espinosa, M.~Quir\'os and C.~Wagner, \plB{355}{95}{209};\\
M.~Carena, M.~Quir\'os and C.~Wagner, \npB{461}{407};\\
H.~Haber, R.~Hempfling and A.~Hoang, \zp{C 75}{97}{539}.

\bibitem{mhiggsEffPot} 
R.~Hempfling and A.~Hoang, \plB{331}{94}{99};\\
R.-J.~Zhang, MADPH-98-1072, hep-ph/9808299.

\bibitem{mhiggs2la}
S.~Heinemeyer, W.~Hollik, G.~Weiglein, \prD{58}{98}{091701}.

\bibitem{mhiggs2lb}
S.~Heinemeyer, W.~Hollik and G.~Weiglein, \plB{440}{98}{296};
KA-TP-17-1998, hep-ph/9812472, to appear in {\sl Eur. Phys. J.} {\bf C}.

\bibitem{ScharfDipl}
R.~Scharf, Diploma Thesis, (Univ.\ of W\"urzburg, 1991).

\bibitem{dreg} 
C.~Bollini, J.~Giambiagi, \nc{B 12}{72}{20};  \\
J.~Ashmore, \ncl{4}{72}{289}.

\bibitem{ga5HV}
G.~'t Hooft, M.~Veltman, \npB{44}{72}{189}.

\bibitem{dred} 
W.~Siegel, \plB{84}{79}{193};\\
D.~Capper, D.~Jones and P.~van Nieuwenhuizen, \npB{167}{80}{479}.

\bibitem{jackjones}
I.~Jack and D.~Jones, hep-ph/9707278, in {\it Perspectives on
Supersymmetry}, ed.~G.~Kane (World Scientific, Singapore), p.~149.

\bibitem{ga5BM}
P.~Breitenlohner and D.~Maison, \cmp{52}{77}{11}.

\bibitem{diffren}
F.~del Aguila, these proceedings;\\
T.~Hahn, M.~P\'erez-Victoria, UG-FT-87-98, hep-ph/9807565.

\bibitem{sbaugw1}
S.~Bauberger and G.~Weiglein, \nim{A 389}{97}{318}.

\bibitem{fea}
J.~K\"ublbeck, M.~B\"ohm and A.~Denner, \cpc{60}{90}{165};\\
H.~Eck, J.~K\"ublbeck, {\it Guide to} \fea {\it 1.0\/}
(Univ.\ of W\"urzburg, 1992);\\
H.~Eck, {\it Guide to} \fea {\it 2.0\/}
(Univ.\ of W\"urzburg, 1995).

\bibitem{two}
G.~Weiglein, R.~Scharf and M.~B\"ohm, \npB{416}{94}{606};\\
G.~Weiglein, R.~Mertig, R.~Scharf and M.~B\"ohm, in {\it New
Computing Techniques in Physics Research 2},
ed.~D.~Perret-Gallix (World Scientific, Singapore, 1992), p.~617.

\bibitem{intnum}
S.~Bauberger, F.A.~Berends, M.~B\"{o}hm and M.~Buza,
\npB{434}{95}{383};\\
S.~Bauberger, F.A.~Berends, M.~B\"{o}hm, M.~Buza and G.~Weiglein,
\nphbps{37}{94}{95}, hep-ph/9406404;\\
S.~Bauberger and M.~B\"{o}hm, \npB{445}{95}{25}.

\bibitem{sirlin}
A.\ Sirlin, \prD{22}{80}{971};\\
W.\ Marciano and A.\ Sirlin, \prD{22}{80}{2695}.

\bibitem{gsw}
P.~Gambino, A.~Sirlin and G.~Weiglein, in preparation.

\bibitem{hhwprec}
S.~Heinemeyer, W.~Hollik and G.~Weiglein, in preparation.

\bibitem{FeynHiggs}
S.~Heinemeyer, W.~Hollik and G.~Weiglein, KA-TP-16-1998,
hep-ph/9812320.

\bibitem{delrhoexp} G.~Altarelli, hep-ph/9811456.

\end{thebibliography}
\end{document}